# Use of Instrumental Neutron Activation Analysis to investigate the distribution of trace elements among subsamples of solid materials

G. D'Agostino[(1)], L. Bergamaschi[(1)], L. Giordani[(1)], M. Oddone[(2)], H. Kipphardt[(3)], S. Richter[(3)]

(1) Istituto Nazionale di Ricerca Metrologica (INRIM) - Unit of Radiochemistry and Spectroscopy, c/o Department of Chemistry, University of Pavia, via Taramelli 12, 27100 Pavia, Italy

(2) Department of Chemistry – Radiochemistry Area, University of Pavia, via Taramelli 12, 27100 Pavia, Italy

(3) Bundesanstalt für Materialforshung und –prüfung (BAM), Richard-Willstätter-Str. 11, 12489 Berlin, Germany

Email: G. D'Agostino, g.dagostino@inrim.it

**Abstract:** The results of analytical measurements performed with solid-sampling techniques are affected by the distribution of the analytes within the matrix. The effect becomes significant in case of determination of trace elements in small subsamples. In this framework we propose a measurement model based on Instrumental Neutron Activation Analysis to determine the relative variability of the amount of an analyte among subsamples of a material. The measurement uncertainty is evaluated and includes the counting statistics, the full-energy gamma peak efficiency and the spatial gradient of the neutron flux at the irradiation position. The data we obtained in a neutron activation experiment and showing the relative variability of As, Au, Ir, Sb and W among subsamples of a highly pure Rh foil are also presented.

## 1. Introduction

In analytical chemistry most of the measurements carried out with solid-sampling techniques are based on ratio methods which compare signals generated by standard and unknown subsamples.

The amount of mass contributing to the net signal usually ranges from a few nanograms to grams, depending on the adopted chemical method. There is experimental evidence that the smaller the subsample mass the larger the relative variability of the amount of analyte among the subsamples. Consequently, a single analytical result will give limited information on the amount of analyte in the material.

Moreover, in trace element analysis the determinations may be not normally distributed and skew for sufficiently small subsample mass. In such cases, Poisson statistics must be considered instead of Gaussian statistics to construct confidence intervals of the results [1].

To some extent, any report which states the elemental composition of a material should include the expected relative standard deviation of the amount of analytes among equal subsamples, at least for the analytes of interest. The information becomes essential both in intercomparisons of small solid-sampling analysis methods [2] and in the development of substances to provide Reference Materials [3] or SI-traceable standards for element determination [4].

In case of well-mixed granular materials, Ingamells suggested to use a sampling constant, $K_s$, to dial with the variability of the amount of analytes among subsamples [1]. In particular, $K_s$ defines the subsample mass (usually in mg) which sets the relative standard deviation of the amount of analyte among subsamples to 1%. In addition to the sampling constant, Kurfurst defined a homogeneity factor, $H_e$, to quantify the degree of homogeneity the analyte in the granular material [5]. The value of $H_e$ estimates the relative standard deviation (usually in percent) of the amount of analyte among subsamples having 1 mg mass.

The most common way to determine the distribution of the amount of analyte among subsamples is to perform absolute determinations of the analyte using the same method under controlled conditions. Since the observed variability includes the instrumental component, the variance of the amount of analyte is estimated by subtraction of the observed variance from the instrumental variance [6]. The computation of the relative standard deviation of the amount of analyte with respect to the mass of the subsample is then trivial. Best results are evidently obtained when the instrumental variability is negligible compared to the observed variability.

In this framework, the evaluation of the contribution to the uncertainty due to the analytical method is compulsory to extract the information concerning the variability of the amount of analyte. This limits the analytical techniques to those which allow a full evaluation of the uncertainty. Among them, Instrumental Neutron Activation Analysis (INAA) has been widely applied because it offers a high metrological quality together with well-known sources of uncertainty [7, 8]

To investigate the full potentiality of INAA in this field, we adapted the neutron activation model to quantify and to evaluate the uncertainty of the ratio between the mass fraction of an analyte in a subsample and the mass fraction of the same analyte in a reference subsample. This allows a direct measurement of the relative variability of the amount of analyte among a set of subsamples and opens a straightforward way to assess if the observed variability is due to the analyte or it is instrumental. The measurement model was used in an experiment intended to determine the distribution of trace elements among subsamples of a highly pure Rh foil.

In this paper we describe the measurement model and the evaluation of the uncertainty contributions. The experimental activity and the results achieved with analysis of the Rh foil are also reported.

**2. Measurement model**

The neutron activation analysis of a sample is based on its exposure to a neutron flux for the production of gamma emitting radionuclides. Since the production and decay paths depend on the irradiated nucleus, the amount of a nuclide within the sample can be quantified by performing a gamma spectrometric measurement.

In particular, the number of target nuclei $N$ of a nuclide can be estimated according to

$$N = \frac{n_c}{(1-e^{-\lambda t_i}) e^{-\lambda t_d} (1-e^{-\lambda t_c})} \frac{\lambda}{k_{ss} R k_{sa} \Gamma k_g \varepsilon} \frac{t_c}{t_{live}}, \qquad (1)$$

where $n_c = n_t - n_b$ is the net count of the full-energy gamma peak, i.e. the difference between the total count and the background count, $R$ is the reaction rate per target nucleus, $\lambda$ is the decay constant, $\Gamma$ is the gamma radiation yield, $t_i$, $t_d$, $t_c$ and $t_{live}$ are the irradiation, decay, counting and live times, $k_{ss}$, $k_{sa}$ and $k_g$ are the irradiation self-shielding, the emission self-absorption and geometry factors and $\varepsilon$ is the detection full-energy gamma peak efficiency for a point-like source located at the center of mass of the sample.

Details about INAA can be found in literature (see for example [9, 10]), however, it is worth to recall that $t_{live} = t_c - t_{dead}$, $\lambda = \ln(2)/t_{1/2}$ and $R = \int \sigma(E) \varphi(E) dE$, where $t_{dead}$ is the dead time of the detection system, $t_{1/2}$ is the half-life of the radionuclide, $E$ is the neutron energy, $\sigma(E)$ is the reaction cross section and $\varphi(E)$ is the energy distribution of the neutron flux.

The ratio between the number of target nuclei of the nuclide in two samples, derived from (1), is

$$\frac{N_1}{N_2} = \frac{\left.\dfrac{n_c}{(1-e^{-\lambda t_i})\,e^{-\lambda t_d}\,(1-e^{-\lambda t_c})}\,\dfrac{\lambda}{k_{ss}\,R\,k_{sa}\,\Gamma\,k_g\,\varepsilon}\,\dfrac{t_c}{t_{live}}\right|_1}{\left.\dfrac{n_c}{(1-e^{-\lambda t_i})\,e^{-\lambda t_d}\,(1-e^{-\lambda t_c})}\,\dfrac{\lambda}{k_{ss}\,R\,k_{sa}\,\Gamma\,k_g\,\varepsilon}\,\dfrac{t_c}{t_{live}}\right|_2},\qquad(2)$$

where the subscripts 1 and 2 refer to sample 1 and sample 2, respectively.

The co-irradiation of the samples and the detection of the same gamma energy peak make the result independent on the knowledge of $t_i$, $\lambda$ and $\Gamma$. Furthermore, in case of spatial gradient of the neutron flux, we can assume with negligible uncertainty that $\varphi_2(E) = \beta_{2,1}\,\varphi_1(E)$, i.e. that the energy distribution of the neutrons irradiating the samples remains constant while the intensity could scale. It follows that the ratio $N_1/N_2$ becomes also independent on the knowledge of $\sigma(E)$. When the samples have the same matrix, shape and dimensions, $k_{ss1} = k_{ss2}$, $k_{sa1} = k_{sa2}$ and $k_{g1} = k_{g2}$. Similarly, $\varepsilon_1 = \varepsilon_2$ if the gamma counting is performed at the same position with respect to the detector. However, in the latter case, it is worth to evaluate the uncertainty due to variations of the counting position.

Equation (2) simplifies to

$$\frac{N_1}{N_2} = \beta_{2,1}\,\frac{\left.\dfrac{n_c}{e^{-\lambda t_d}\,(1-e^{-\lambda t_c})}\,\dfrac{t_c}{t_{live}}\right|_1}{\left.\dfrac{n_c}{e^{-\lambda t_d}\,(1-e^{-\lambda t_c})}\,\dfrac{t_c}{t_{live}}\right|_2}.\qquad(3)$$

Timing measurements, i.e. $t_d$, $t_c$ and $t_{live}$, can be usually performed with negligible uncertainty. Thus, if the non-linearity of the model and the correlation among the parameters are neglected, the relative uncertainty of the ratio is

$$u_r\!\left(\frac{N_1}{N_2}\right) = \sqrt{u_r^2(n_{c1}) + u_r^2(n_{c2}) + u_r^2(\varepsilon_1) + u_r^2(\varepsilon_2) + u_r^2(\beta_{2,1})},\qquad(4)$$

where the subscript r means relative.

Since the information concerning the scaling factor $\beta_{2,1}$ is often unavailable, it is of common practice to assume $\beta_{2,1} = 1$ and to add the contribution to the uncertainty based on previous knowledge of the spatial gradient of the neutron flux at the irradiation position. However, equation (3) shows that a direct measurement of $\beta_{2,1}$ is possible when the amount of a nuclide activated during the irradiation is known and can be used as an internal monitor of the flux [11]. In this case, the correction factor can be measured according to

$$\beta_{2,1} = \frac{\left.\dfrac{n_c}{e^{-\lambda t_d}(1-e^{-\lambda t_c})}\dfrac{t_c}{t_{live}}\right|_2}{\left.\dfrac{n_c}{e^{-\lambda t_d}(1-e^{-\lambda t_c})}\dfrac{t_c}{t_{live}}\right|_1} \frac{N_1}{N_2} . \tag{5}$$

The relative uncertainty is

$$u_r(\beta_{2,1}) = \sqrt{u_r^2(n_{c1}) + u_r^2(n_{c2}) + u_r^2(\varepsilon_1) + u_r^2(\varepsilon_2)} . \tag{6}$$

Let us now consider a set of *n* subsamples of a material and having a similar mass *m*. If the isotopic abundance of the target nuclide of an analyte is constant, (3) can be used to compute the ratio, $\alpha$, between the mass fraction of the analyte in sample i, $w_i$, and the mass fraction of the same analyte in a reference sample, $w_{ref}$, according to

$$\alpha_{i,ref} = \frac{w_i}{w_{ref}} = \frac{N_i/m_i}{N_{ref}/m_{ref}} . \tag{7}$$

If the masses of the subsamples are known with negligible uncertainty, the relative uncertainty is

$$u_r(\alpha_{i,ref}) = \sqrt{u_r^2(n_{c1}) + u_r^2(n_{cref}) + u_r^2(\varepsilon_1) + u_r^2(\varepsilon_{ref}) + u_r^2(\beta_{ref,i})} . \tag{8}$$

The distribution of the *n*-1 ratios $\alpha$ is a direct measure of the relative variability of the analyte among the subsamples. Moreover, the relative standard deviation of the ratios $\alpha$ corresponds to the relative standard deviation of the amount of analyte among subsamples. It is worth to notice that the result is not dependent on the choice of the reference sample.

## 3. Evaluation of the uncertainty contributions

Equation (8) shows that the counting statistics, the full-energy peak efficiency of the detection system and the spatial gradient of the neutron flux at the irradiation position are the uncertainty contributions of $\alpha$.

The contribution due to counting statistics can be computed taking into account that the number of collected counts have a Poisson distribution. It can be shown that the relative variance of $n_c$ is

$$u_r^2(n_c) = \frac{n_c + 2 n_b}{n_c^2}, \qquad (9)$$

where $n_c$ is the net count and $n_b$ is the background count at the full-energy peak. Best results are obtained when $n_c$ is the maximum allowed by the experimental conditions and $n_b$ is negligible.

A first way out to increase $n_c$ with respect to $n_b$ is to decrease the distance between the sample and the detector. However, there are drawbacks such as the effects of possible variations of the distance and of true coincidence summing. To limit both the effects, the gamma counting is never performed in proximity of the detector. In this condition the full-energy peak efficiency can be approximated with

$$\varepsilon = \frac{k}{d^2}, \qquad (10)$$

where $k$ depends mainly on the characteristics of the detection system, such as dimensions of the crystal and the peak-to-Compton ratio, and $d$ is the distance between the center of mass of the sample and the imaginary point inside the detector where the efficiency would tend to infinite. Actually, the distance $d$ depends on the energy of the detected gamma photon. Nevertheless, for the evaluation of the uncertainty, we consider an average $d$ value and we neglect the effect of the approximation.

From (10) it can be shown that a finite variation of the distance, $\Delta d$, which occurs at a distance $d \gg \Delta d$, corresponds to a relative variation of the efficiency according to

$$\frac{\Delta \varepsilon}{\varepsilon} = -2 \frac{\Delta d}{d}. \qquad (11)$$

For example, let us consider cylindrical samples having height $h$ and fixed to a sample holder with the axis perpendicular to the entrance surface of the detector (Figure 1). A departure of the sample holder from the nominal position, $\Delta p$, and a deviation of the height of the sample from the nominal value, $\Delta h$, sum up to give a variation $\Delta d$.

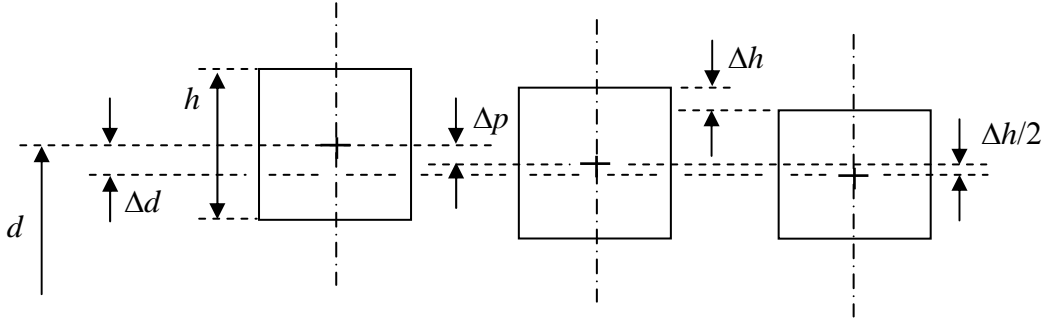

**Figure 1:** The variation of the distance $\Delta d$ due to the variations of the positioning system $\Delta p$ and of the sample height $\Delta h$.

Under the hypothesis of a uniform distribution both for $\Delta p$ and $\Delta h$, from (11) it follows that the relative variance of $\varepsilon$ is

$$u_r^2(\varepsilon) = \frac{4}{d^2}\left(\frac{(\Delta p)^2}{3} + \frac{(\Delta h/2)^2}{3}\right). \tag{12}$$

Finally, the contribution to the uncertainty due to the spatial gradient of the neutron flux at the irradiation position can be evaluated according to (6) in case of use of (5) to correct the effect. Conversely, when the correction is not possible, the contribution should be evaluated taking into account a previous knowledge of the gradient of neutron flux at the irradiation position.

## 4. Experimental

We used model (7) in an experiment based on INAA to measure the relative variability of the amount of contaminant elements among subsamples of a highly purified Rh foil.

The main neutron capture reactions occurring in Rh are $^{103}$Rh(n, $\gamma$)$^{104m}$Rh ($t_{1/2}$ = 4.34 min), $^{103}$Rh(n, $\gamma$)$^{104}$Rh ($t_{1/2}$ = 42.3 s), $^{103}$Rh(n, p)$^{103}$Ru ($t_{1/2}$ = 40 d), $^{103}$Rh(n, 2n)$^{102m}$Rh ($t_{1/2}$ = 208 d) and $^{103}$Rh(n, 2n)$^{102}$Rh ($t_{1/2}$ = 207 d). Since the reaction cross sections for the production of $^{103}$Ru via (n, p), $^{102}$Rh and $^{102m}$Rh via (n, 2n) are very low, the activity due to the matrix is negligible after about one our from the end of the irradiation. This makes Rh a good material to be investigated by INAA. Moreover, the production of gamma emitters from the matrix allows the internal monitoring of the neutron flux. The correction factor of the flux can be computed with (5) because the ratio between the number of $^{103}$Rh within the subsamples, $N_1/N_2$, corresponds to the ratio between their (known) masses, $m_1/m_2$.

A set of twelve circular subsamples (6 mm diameter, 25 µm thickness and 9 mg mass) was obtained by laser cutting the foil (Figure 2).

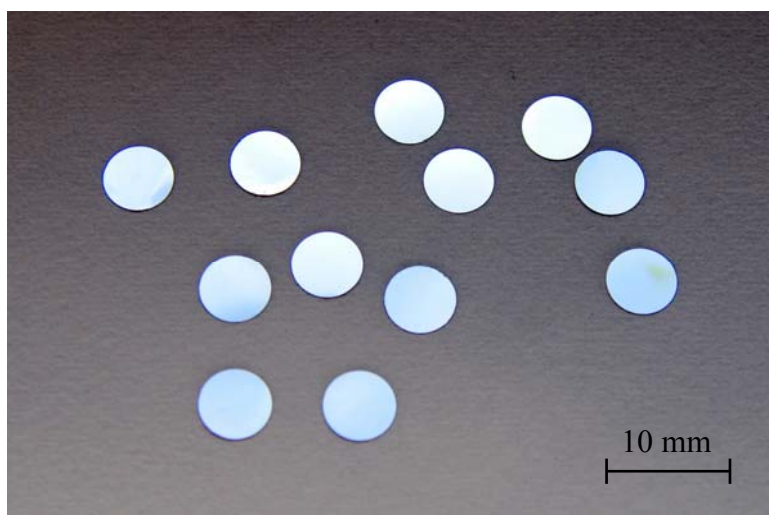

**Figure 2:** The subsamples of the Rh foil

The masses of the subsamples were measured with a digital analytical balance which was calibrated using weights with SI traceable values and having a resolution of 0.01 mg. Next, the subsamples were sealed in polyethylene vials (cylindrical shape, 7 mm diameter and 9 mm height), packed in aluminum containers and co-irradiated at the irradiation facility. To minimize the effect of spatial gradients and depression of the neutron flux, the twelve subsamples were divided in two groups and stacked on two planes at a distance of about 1 cm (Figure 3).

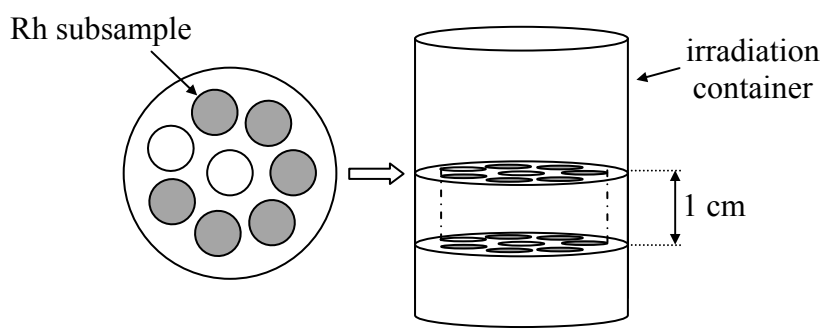

**Figure 3:** Position of the Rh subsamples in the irradiation container

The neutron irradiation lasted 7 h and was performed in the central thimble of the 250 kW TRIGA Mark II reactor at the Laboratory of Applied Nuclear Energy (LENA) of the University of Pavia. The thermal and epithermal neutron fluxes were about $6 \times 10^{12}$ cm$^{-2}$ s$^{-1}$ and $5.5 \times 10^{11}$ cm$^{-2}$ s$^{-1}$, respectively.

At the end of the irradiation, the subsamples were (i) left to cool until the activity decayed to a safe value, (ii) taken away from their vials and rinsed with diluted nitric acid and pure water in order to remove any contamination occurred during preparation and handling and finally (iii) put in polycarbonate containers for counting. To assure the greatest reproducibility of the position during the gamma counting each subsample was directly fixed at the center of the bottom surface of its container. The bottom of the counting container was machined in order to limit the deviation of the thickness from the average value below a tenth of a millimeter.

Gamma spectrometry was performed with an automatic system including a sample changer, a coaxial Germanium detector Canberra® GC3518 (relative efficiency 35%, resolution 1.8 keV

FWHM at 1332 keV), a digital signal processor ORTEC® DSPEC, and a personal computer running a software for data acquisition and processing ORTEC® Gamma Vision 6.0.

The position of the sample holder with respect to the distance *d* was measured using the method described in [12]. During the counting, we kept the position of the subsamples to a distance *d* of about 10 cm. The gamma counting of the twelve subsamples was repeated in different periods and with different counting times. In particular the twelve spectra were recorded six times in a sequence of measurement runs performed 20 h, 1 d, 2 d, 6 d, 9 d and 20 d after the end of the irradiation with counting times of 15 min, 1 h, 2 h, 3 h, 12 h and 24 h, respectively.

## 5. Results

The instrumental neutron activation analysis indicated traces of As, Au, Cu, Ir, La, Na, Pd, Pt, Sb and W. However, to obtain reliable results, we rejected the outliers in repeated measurement runs and the data affected by significant departures of the full-energy gamma peak from the Gaussian shape. This excluded from the analysis Cu, La, Na, Pd and Pt and limited the useful measurement runs for the remaining analytes.

The repeated measurement runs used to compute the ratios $\alpha$ of the quantified analytes are shown in Table 1. It is also reported the nuclear reaction, the half-life of the produced radionuclide and the energy of the adopted gamma peak.

| Analyte | Meas. runs | Nuclear reaction | $t_{1/2}$ | Energy peak / keV |
|---------|------------|------------------|-----------|-------------------|
| As | 1, 2, 3 | $^{75}$As(n, γ)$^{76}$As | 25.9 h | 559.1 |
| Au | 2, 3, 4 | $^{197}$Au(n, γ)$^{198}$Au | 2.70 d | 411.8 |
| Ir | 1, 2, 3, 4, 5, 6 | $^{191}$Ir(n, γ)$^{192}$Ir | 73.8 d | 316.5 |
| Sb | 5 | $^{121}$Sb(n, γ)$^{122}$Sb | 2.73 d | 564.1 |
| W | 1, 2, 3, 4 | $^{186}$W(n, γ)$^{187}$W | 23.7 h | 479.5 |

**Table 1:** The measurement runs, the nuclear reactions, the half-life of the radionuclides and the energy of the gamma peak used to compute the ratios $\alpha$ of the quantified analytes.

Since the production and gamma emission at 475.1 keV of $^{102}$Rh and $^{102m}$Rh are free of interference, we computed the correction factors of the flux according to (5) and (6) taking subsample 1 as the reference and by recording the gamma peak during the measurement runs 4, 5 and 6. The values, reported in Figure 4, are the averages of the results obtained in the repeated runs and weighted by their uncertainties. The weighted uncertainties of the average values resulted to be about 1 %. With the exception of the value $\beta_{Rh1,Rh4}$, the results show a difference of about 4 % of the neutron flux intensities between the two planes where the subsamples were irradiated.

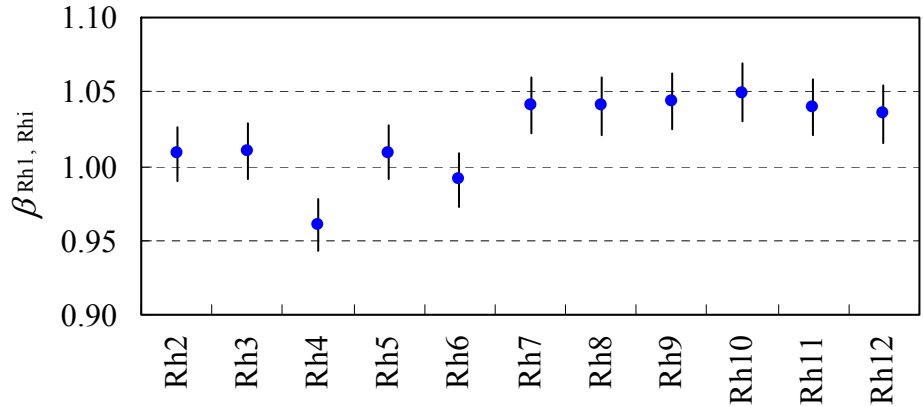

**Figure 4:** The correction factors of the gradient of the neutron flux which irradiated the Rh subsamples (interval bars with a coverage factor 2).

The ratios $\alpha$ were computed for each measurement run according to (7) and (8), taking subsample 1 as the reference and correcting the effect due to the gradient of the neutron flux at the irradiation position.

The contribution to the uncertainty due to the variation of the full-energy peak efficiency was evaluated according to (12) by assuming a maximum departure of the sample holder from the nominal position $\Delta p = 0.5$ mm and a negligible deviation of the thickness of the Rh foil from the nominal value, $\Delta h \cong 0$ mm. The measurement repeatability of the detection system was tested by counting in repeated runs a multi-gamma source located at a distance $d$ of about 10 cm. The best results showed a relative standard deviation of the net count of the full-energy gamma peak, $n_c$, of 0.6 %, which was in good agreement with 0.5 %, the value obtained according to (9) and evaluating the contribution to the uncertainty due to counting statistics.

The graphs in Figure 5 show the averages of the ratios $\alpha$ obtained in repeated runs and weighted by their uncertainties.

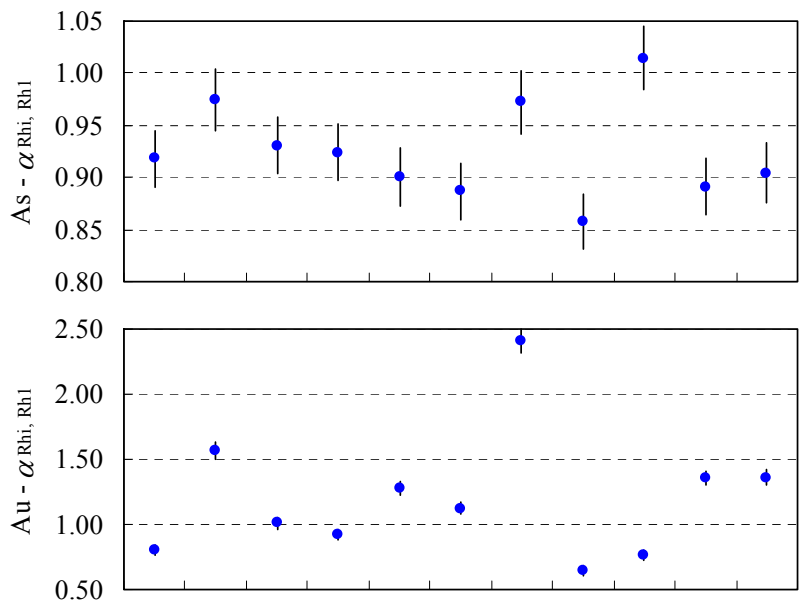

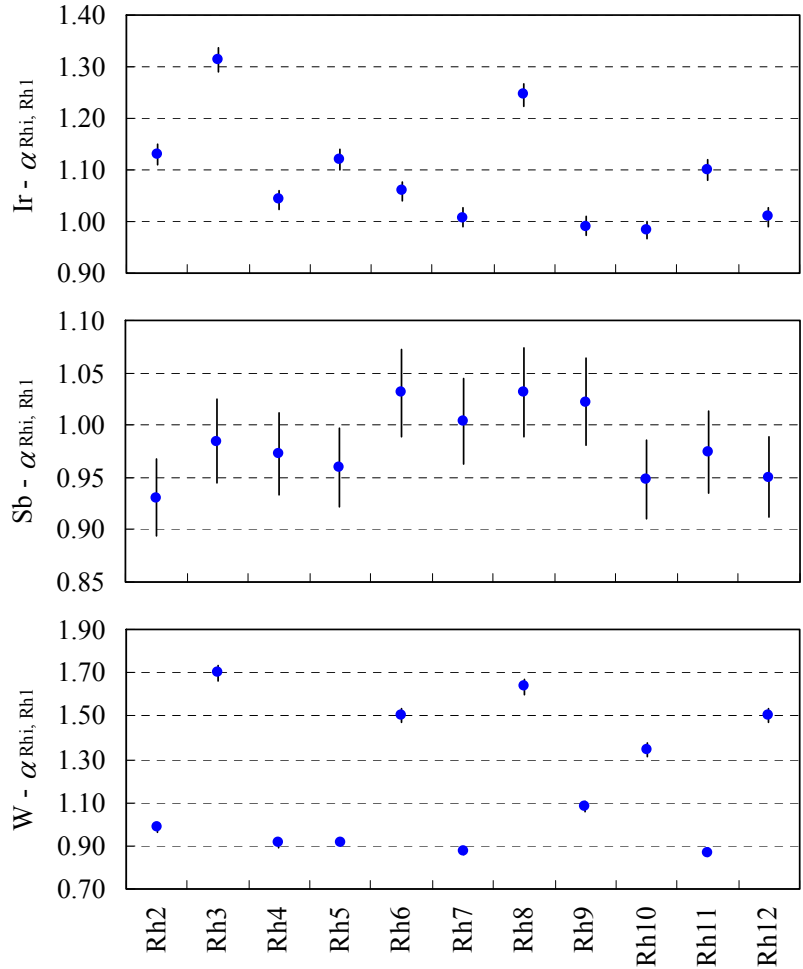

**Figure 5:** The measured relative variability of the amount of As, Au, Ir, Sb and W among the Rh subsamples (interval bars with a coverage factor 2).

The weighted uncertainties (coverage factor 2) of the averaged ratios $\alpha$ of As, Au, Ir, Sb and W are about 2.4 %, 4 %, 1.2 %, 2.4 % and 1.4 %, respectively. According the results of the analysis, the relative standard deviations of the amount of As, Au, Ir, Sb and W among the subsamples of the Rh foil are 5 %, 40 %, 10 %, 3.5 % and 27 %, respectively.

The statistical analysis of the variability of the amount of quantified analytes is beyond the scope of the present work. However, it is worth to point out that the amount of Au in subsample Rh8 is roughly 5 times the amount of Au in subsample Rh9. In addition, the variability of the ratios $\alpha$ of Au and Ir show the possible asymmetry in repeated measurements of a trace-element in subsamples of a highly purified materials [1].

## 6. Conclusions

We explored the potentiality of INAA in analytical chemistry to measure the relative variability of trace elements among subsamples of solid materials by proposing a model based on the neutron activation equation to determine the ratio between the mass fractions of an analyte in two samples. The measurement uncertainty was fully evaluated.

To test the model, we designed and carried out an experiment to find out the distribution of contaminant elements among subsamples of a highly pure Rh foil. The measurement quantified the relative variability of As, Au, Ir, Sb and W with a relative uncertainty (coverage factor 2) ranging between 1.2 % and 4 %.

**Acknowledgments**

This work was jointly funded by the European Metrology Research Programme (EMRP) participating countries within the European Association of National Metrology Institutes (EURAMET) and the European Union. The authors would like to thank J. Noordmann for making available the Rh foil and laser cutting of the subsamples. Moreover, they are grateful to M. Santiano for the machining of the irradiation containers and to F. Pennecchi for fruitful discussions concerning the statistical analysis of the data.